# Boson metastable ground states with spontaneous symmetry breaking

M. Chaves [1]

*Escuela de Fisica*
*Universidad de Costa Rica*
*Ciudad Universitaria Rodrigo Facio*
*San Jose, Costa Rica*

We show that a system of bosons in a $T = 0$ quantum field theory can present metastable ground states with spontaneous symmetry breaking, even in the absence of an imaginary mass term. This gives a natural explanation to the Davis-Shellard background field $e^{-i\omega_0 t}$ and adds a new degree of freedom in boson systems, with possible applications in cosmology, condensed matter and high energy physics.

[1] FAX: (506) 207-5619; Office tel. (506) 207-5394; Email: mchaves@cariari.ucr.ac.cr

Quantum field theories with boson fields $\varphi(\mathbf{x},t)$ are nowadays ubiquitous in theoretical physics. Often used in condensed matter, cosmology and high energy theories, they are particularly useful in situations where it is important that there exists spontaneous symmetry breaking (SSB).[1] In this Letter we show that some boson systems can present metastable ground states where SSB is present, in situations where this is not usually considered possible.

A few years ago it was proved[2] that a special kind of background field $e^{-i\omega_0 t}$ has to exist in order for cosmic strings[3] and strings in a complex boson field[4] to possess the physical qualities usually associated with vortices. It is necessary in order for cosmic strings to carry angular momentum, or for a string to obey the Magnus force law of fluid vortices. Since then it has been used in the study of vortices in superconductors,[5] electroweak vortices,[6] cosmology[7] and Galilean extended objects.[8] This field is a byproduct of the metastable ground states explained here.

It was noticed[9] using second-quantization methods that Bose-Einstein condensation in a system of complex bosonic particles results in massless pseudoparticles. If a boson quantum field theory Lagrangian has a mass term with the customary sign, we say that it has a <u>R</u>eal <u>M</u>ass. An example is the following Lagrangian:

$$\mathcal{L}_{\text{RM}} = \left|\frac{\partial \varphi}{\partial t}\right|^2 - |\nabla\varphi|^2 - m^2|\varphi|^2 - \lambda|\varphi|^4 \quad , \tag{1}$$

$\lambda > 0$. Observe it has a global phase $U(1)$ symmetry $\varphi \to e^{i\alpha}\varphi$, $\alpha$ a constant. The Hamiltonian is

$$\mathcal{H}_{\text{RM}} = |\pi|^2 + |\nabla\varphi|^2 + m^2|\varphi|^2 + \lambda|\varphi|^4 \quad , \quad \pi = \frac{\partial}{\partial\dot{\varphi}}\mathcal{L}_{\text{RM}} = \dot{\varphi}^* \quad , \tag{2}$$

from which it is evident that the ground state of the system is $\varphi = 0$. The energy term $\lambda|\varphi|^4$ represents a repulsive force (increasing the number density $|\varphi|^2$ increases the energy), while the mass term acts effectively like a repulsive force, too, for the same reason.

A system frequently used to study SSB of a global symmetry is given by the <u>I</u>maginary <u>M</u>ass Lagrangian

$$\mathcal{L}_{IM} = \left|\frac{\partial \varphi}{\partial t}\right|^2 - |\nabla\varphi|^2 + 2\lambda\eta^2|\varphi|^2 - \lambda|\varphi|^4, \tag{3}$$

$\eta$ being real, which is exactly the same as (1) except for the sign of the mass term and a convenient choice for the algebraic form of this term. Its Hamiltonian is

$$\mathcal{H}_{IM} = \left|\frac{\partial \varphi}{\partial t}\right|^2 + |\nabla\varphi|^2 + \lambda\left(|\varphi|^2 - \eta^2\right)^2 - \lambda\eta^4, \tag{4}$$

from which it is clear, using the effective potential method,[10] that the ground state of the system occurs for $\varphi = \eta$. This fixing of $\varphi$'s phase constitutes the SSB. The field square term now represents an attractive force. If we were to view it as a mass term, from the identification $m^2 = -2\lambda\eta^2$ it would follow that the mass is imaginary, hence the name. The SSB of the $U(1)$ symmetry should result in an massless pseudoparticle, according to Goldstone theorem[1] and in accord with the mentioned result.[9]

Is it possible for a system with a real mass term to undergo SSB? The grand canonical distribution of an aggregate of conserved bosons[11] has a chemical potential that acts like an imaginary mass, so that it can generate SSB. Feynman's path integral formulation of the partition function[12] was applied to quantum field theory, resulting in its finite-temperature formulation. This setup was used extensively to study the temperature dependence of both the ground state of an aggregate of self-interacting bosons and of the Higgs bosons of grand unified theories.[13] The chemical potential enters as an imaginary mass term that can produce SSB if the temperature is cold enough.[14]

Here we show there is a way to obtain an imaginary mass in a system that does not originally contain one, in the context of a $T = 0$ quantum field theory. As a result there exists a class of metastable ground state solutions of Lagrangian (1) that can undergo SSB and possess a background field $e^{-i\omega_0 t}$, precisely as postulated by Davis and Shellard.[2] Loosely speaking, it is set up by the system in situations when a forming a condensate is necessary but impossible otherwise.

We proceed by transforming to new fields that make the metastable states more evident. We first perform a contact transformation between the original canonical fields $(\varphi, \pi)$ of (1) and new fields $(\psi, \Pi)$. (More specifically, it is a point transformation.) The generating function is taken to be

$$\mathcal{F} = e^{i\omega_0 t}\Pi\varphi + e^{-i\omega_0 t}\varphi^*\Pi^* , \qquad (5)$$

from which, employing the canonical methods of contact transformations, we obtain the algebraic relations between the old and new fields:

$$\pi = \tfrac{\partial}{\partial \varphi}\mathcal{F} = e^{i\omega_0 t}\Pi, \qquad \pi^* = \tfrac{\partial}{\partial \varphi^*}\mathcal{F} = e^{-i\omega_0 t}\Pi^* , \qquad (6)$$

and

$$\psi = \tfrac{\partial}{\partial \Pi}\mathcal{F} = e^{i\omega_0 t}\varphi, \qquad \psi^* = \tfrac{\partial}{\partial \Pi^*}\mathcal{F} = e^{-i\omega_0 t}\varphi^* . \qquad (7)$$

One should be aware that for $\omega_0 = 0$ the identity transformation is generated, and that, evidently, $\Pi = \dot{\psi}^* + i\omega_0\psi^*$. From

$$\mathcal{H}' = \mathcal{H} + \tfrac{\partial}{\partial t}\mathcal{F} \qquad (8)$$

and the usual canonical methods it is possible to find the new Lagrangian and Hamiltonian:

$$\mathcal{L}' = \left|\tfrac{\partial \psi}{\partial t}\right|^2 + i\omega_0\psi^*\overleftrightarrow{\partial_t}\psi - |\nabla\psi|^2 + (\omega_0^2 - m^2)|\psi|^2 - \lambda|\psi|^4 \qquad (9)$$

and

$$\mathcal{H}' = \Pi^*\Pi + i\omega_0\Pi\psi - i\omega_0\psi^*\Pi^* + |\nabla\psi|^2 + m^2|\psi|^2 + \lambda|\psi|^4. \qquad (10)$$

We have finish describing the system using the new fields.

Now, the quantum mechanical amplitude $A$ between an initial and a final state can be written, using functional methods, in the form:

$$A = \int D\varphi D\varphi^* D\pi D\pi^* \exp[i\int dt \int d^3x (\pi\dot{\varphi} + \dot{\varphi}^*\pi^* - \mathcal{H})]. \qquad (11)$$

From the definition of a contact transformation it follows that

$$\pi \dot{\varphi} + \dot{\varphi}^* \pi^* - \mathcal{H} = \Pi \dot{\psi} + \dot{\psi}^* \Pi^* - \mathcal{H}' + \frac{d}{dt}\mathcal{F}. \tag{12}$$

(Observe in (8) the time derivative is partial and in (12) is total.) This equation implies that the functional integral

$$\int D\psi D\psi^* D\Pi D\Pi^* \exp[i \int dt \int d^3x (\Pi \dot{\psi} + \dot{\psi}^* \Pi^* - \mathcal{H}')] \tag{13}$$

differs from $A$ by just a phase $\exp\{\int d^3x [\mathcal{F}(t_f) - \mathcal{F}(t_i)]\}$ that depends only on the endvalues. Here $t_i$ and $t_f$ are the initial and final times. If $\mathcal{F}$ has the same values initially and finally then the amplitude does not depend at all on which pair of canonical variables we use. Furthermore, any expected values calculated using either (11) or (13) will give the same result independently of the endvalues of $\mathcal{F}$, since normalization will cancel the phase.

In order to have SSB (and interesting results) $\omega_0$ has to satisfy $\omega_0 > m$. Assume this to be the case, and define the positive number $\eta$ by $2\lambda\eta^2 \equiv \omega_0^2 - m^2$. With its help the potential implied by $\mathcal{L}'$ can be written in the form

$$V = \lambda|\psi|^4 - 2\lambda\eta^2|\psi|^2 = \lambda\left(|\psi|^2 - \eta^2\right)^2 - \lambda\eta^4, \tag{14}$$

which suggests that the constant and uniform state $\psi = \psi_0 = \eta$ could be a metastable ground state of the system. The Euler-Lagrange equation generated by $\mathcal{L}'$ is

$$\frac{\partial^2 \psi}{\partial t^2} - 2i\omega_0 \frac{\partial \psi}{\partial t} - \nabla^2 \psi + 2\lambda\left(|\psi|^2 - \eta^2/\lambda\right)\psi = 0, \tag{15}$$

after integration by parts (we assume the Lagrangian to be the integrand of an action), and does have the solution $\psi_0$. If we rewrite (10) in terms of $\psi$ and $\dot{\psi}$ we get

$$\mathcal{H}' = \left|\frac{\partial \psi}{\partial t}\right|^2 + |\nabla\psi|^2 + \lambda\left(|\psi|^2 - \eta^2\right)^2 - \lambda\eta^4, \tag{16}$$

which tells us that that $\psi_0$ is a metastable ground state of the system and SSB has to occur, at least as long as $\mathcal{H}'$ represents the energy.

From (6) and (8) one can immediately show that the generating function is numerically related to the particle number:

$$\mathcal{H}' - \mathcal{H} = \frac{\partial}{\partial t}\mathcal{F} = \frac{\partial}{\partial t}\frac{d}{dt}(\varphi^*\varphi). \tag{17}$$

In those cases where $\varphi^*\varphi$ is conserved the state $\psi_0$ will be the ground state and thus stable, as, for example, in systems composed of stable molecules. Even in cases where the particle number is varying, it may be that the state will take a long time to decay to the true vacuum. For example, if $\varphi^*\varphi \propto e^{t/\tau}$, then the effect on $\mathcal{H}'$ will go as $1/\tau^2$, which can be fairly small for large lifetimes. We mentioned that there was a phase difference $\exp\{\int d^3x[\mathcal{F}(t_f) - \mathcal{F}(t_i)]\}$ in the quantum amplitudes calculated with the old or new variables. We see that for systems with a conserved particle number there is no phase difference.

So we come to the conclusion that there are metastable ground states given by $\varphi_0 = e^{-i\omega_0 t}\eta$ if $\int \varphi^*\varphi \, d^3x$ is conserved. A special example is the nonrelativistic regime of Lagrangian (1). In this case $\omega_0 = m$, there is no SSB and one also neglects the energy of the field $\psi$ as compared with $m$. The Lagrangian becomes

$$\mathcal{L}_{NR} = 2im\psi^*\partial_t\psi - |\nabla\psi|^2 - \lambda|\psi|^4, \tag{18}$$

after integration by parts, that has only to be divided by $2m$ to have the usual form. In this case the formalism assures us that it is possible to do nonrelativistic quantum mechanics ignoring the rest mass energy as long as the particle number is conserved and the energies involved are not too large.

The two imaginary mass Lagrangians (3) and (9) differ in the extra term $i\omega_0\psi^*\overleftrightarrow{\partial_t}\psi$ the latter one has, which modifies the nature of the pseudoparticles resulting after the SSB. To determine the spectrum at tree level we substitute $\psi = \eta + (\psi_R + i\psi_I)/\sqrt{2}$ in (3) and (9), then discard terms higher than second-order. For (3) this results in two separated Lagrangians with dispersion relations $\omega_1 = k$ and $\omega_2^2 = k^2 + M^2$, where $M^2 \equiv 4\lambda\eta^2$, so that there are two types of pseudoparticles, one a massless Goldstone that goes *at the speed of light* and the other a massive particle. In the case of (9) the result is

$$\mathcal{L}' = \tfrac{1}{2}\dot{\psi}_R^2 + \tfrac{1}{2}\dot{\psi}_I^2 + \omega_0 \psi_I \overleftrightarrow{\partial}_t \psi_R - \tfrac{1}{2}(\nabla\psi_R)^2 - \tfrac{1}{2}(\nabla\psi_I)^2 - \tfrac{1}{2}M^2\psi_R^2, \tag{19}$$

that has to be diagonalized in order to make physical sense out of it. Similar calculations have been done in the case without SSB.[14] Writing (19) in matrix form in the momentum picture for the two fields $\psi_R$ and $\psi_I$, we see that the dispersion relations are the zeros of the diagonal components *after* diagonalization. But these are obviously the same as the zeroes of the determinant of the matrix *before* diagonalization, so all we have to do is to solve for $\omega$ the equation

$$\det\begin{pmatrix} \omega^2 - k^2 - M^2 & i\omega_0\omega \\ -i\omega_0\omega & \omega^2 - k^2 \end{pmatrix} = 0. \tag{20}$$

There is only one physical solution which, in the limit of small $k$ is a photon $\omega = vk$ where $v$ is a speed given by $v^2 = (\omega_0^2 - m^2)/(3\omega_0^2 - m^2)$, so that it is *less than the speed of light*. Thus the system given by (9) is not Lorentz invariant, in accord with previous ideas.[2,8] The background field fixes the frame of reference.

We have seen how real mass Lagrangians in a $T = 0$ quantum field theory can form metastable ground states with SSB. A Davis-Shellard background is established that changes the sign of the mass term, as if the system lifted itself by its bootstraps. It is gratifying not to have to postulate the Davis-Shellard *ad hoc*, but to obtain it in a natural way from a very ordinary Lagrangian. It is a kind of chemical potential, but instead of being a Lagrange multiplier characterizing the most probable distribution of copies of the system in some ensemble, as the usual chemical potential, it is a parameter in an exact solution of a quantum field theory. Notice these results cannot be applied to real fields. It is the global $U(1)$ invariance in the Lagrangian that allows for the "hiding" of the $e^{-i\omega_0 t}$ in most of its terms so that the system is not changed too much. Nothing similar can be done with sines or cosines.